\documentclass{article}
\usepackage{spconf,amsmath,multirow,booktabs}
\usepackage{graphicx}
\usepackage{comment}
\usepackage{xcolor}
\usepackage{graphicx}
\usepackage{subcaption}
\usepackage[normalem]{ulem}

\title{Improving Speaker Identification for Shared Devices \\by Adapting Embeddings to Speaker Subsets}

\name{Zhenning Tan, Yuguang Yang, Eunjung Han, Andreas Stolcke}
\address{Amazon Alexa AI, Sunnyvale, CA, USA}

\begin{document}
\maketitle

\begin{abstract}
Speaker identification typically involves three stages.
First, a front-end speaker embedding model is trained to embed utterance and speaker profiles. Second, a scoring function is applied between a runtime utterance and each speaker profile. Finally, the speaker is identified using nearest neighbor according to the scoring metric. To better distinguish speakers sharing a device within the same household, we propose a household-adapted nonlinear mapping to a low dimensional space to complement the global scoring metric. The combined scoring function is optimized on labeled or pseudo-labeled speaker utterances. With input dropout, the proposed scoring model reduces EER by {\color{black}45-71\%} in simulated households with 2 to 7 hard-to-discriminate speakers per household. On real-world internal data, the EER reduction is 49.2\%. From t-SNE visualization, we also show that clusters formed by household-adapted speaker embeddings are more compact and uniformly distributed, compared to clusters formed by global embeddings before adaptation. 
\end{abstract}

\section{Introduction}
\vspace{-1em}
Voice-controlled electronic consumer products are becoming more and more popular, allowing users to interact with devices using speech. While personal devices like smartphones or wristbands are primarily operated by a single user, other products, such as Amazon Echo and Google Home, are shared by multiple users, enabled by far-field speech recognition. For such communal devices high-accuracy speaker recognition is critical to distinguish among the multiple users, which typically belong to the same household.
Speaker recognition enables personalized experiences, such as  playing a user's preferred music, or restricting unauthorized access to secured services, such as payment for shopping or access to personal email.
Speaker recognition is used in either closed-set or open-set scenarios. Closed-set speaker identification assumes that all speakers are known from the training set and can therefore be treated as a multi-class classification problem \cite{Li2020}. Open-set speaker identification allows for some speakers to be unseen in training, to be identified as ``guest speakers'', making it more widely applicable \cite{fortuna2005open, angkititrakul2007discriminative, wilkinghoff2020open}.

Speaker recognition typically involves three stages. First, a universal speaker encoder computes utterance-level speaker embeddings from audio (\textit{embedding extraction front-end}). To compute the profile embedding from multiple enrollment utterances, we may L2-normalize and average the utterance embeddings. Second, at the household level, we compare a test utterance embedding to each of the enrolled speaker profile embeddings using a scoring function (\textit{scoring back-end}). Finally, we identify the speaker as the one with highest matching score.
If that score is higher than a predefined threshold we accept the hypothesis that test utterance comes from the associated enrolled speaker. Otherwise, we reject the hypothesis and conclude that test utterance comes from a guest speaker (\textit{hypothesis testing}).
For embedding extraction, i-vectors \cite{dehak2010front}, d-vectors \cite{heigold2016end, wan2018generalized, chung2018voxceleb2}, x-vectors \cite{snyder2017deep, snyder2018x}, and other methods \cite{li2017deep, garcia2020magneto, desplanques2020ecapa} are commonly used. The scoring function can be based on cosine similarity, probabilistic linear discriminant analysis (PLDA)\cite{ioffe2006probabilistic, prince2007probabilistic} and its variants \cite{kenny2010bayesian, ferrer2020speaker}, or a deep neural net \cite{garcia2020magneto, pelecanos2021dr}.

One of the problems with this common approach is that embedding training requires large sets of speakers and the whole system is therefore trained to differentiate \textit{arbitrary} speakers, which could be sub-optimal for speakers in a given household. For example, family members in the same household tend to have similar voice characteristics (e.g., accents) and share similar acoustic environments (e.g., reverberant room), which makes them more challenging to differentiate than arbitrary speakers. To overcome this limitation we could incorporate household-specific information into the front-end embedding extractor or the back-end scoring function. A naive strategy is to fine-tune the front-end embedding extractor using household-specific data to generate household-specific front-end embeddings. However, given the typical size and complexity of the front-end embedding model, the available household-specific data is typically insufficient for fine-tuning. Moreover, even if they could be fine-tuned, it is not scalable to train and store full-sized embedding models for each household.

In this work, we propose a lightweight, scalable household-adapted back-end scoring model to better distinguish household members. Our approach has three steps: first, we use a small neural network to extract household-specific low-dimensional features from universal embeddings; second, we compute local speaker comparison scores from the low-dimensional features; finally, we fuse the local score with the global scores based on universal embeddings to obtain the final scores.
Our approach resembles the idea of adapting a pretrained model to specific tasks in a reduced parameter space, which has been broadly applied in natural language processing, computation vision, and speech domains.  
It is also similar to embedding adaptation in few-shot learning in computer vision\cite{ye2020few}, and intra-conversation variability analysis in speaker diarization \cite{shum2011exploiting}. 

Our approach also works well with on-device household-specific speaker recognition, where all modelling tasks (data acquisition, training, and inference) are completed directly on the device without personal data being sent to a server. Additionally, the household-specific embedding adaptation information could be sent back to the server to further improvement the global embedding extraction model using federated learning \cite{konevcny2016federated}. 

Our main contributions are as follows: We propose a novel light-weight household-adapted scoring model that can learn low-dimensional embeddings optimized for speaker subsets, such as found in household settings. We investigate the effect of pseudo-labeling error on the adaptation process, and propose an effective embedding-level dropout method to make the adapted embeddings more robust to label noise.
Finally, we show that the adapted embeddings form more compact and uniformly distributed clusters. The proposed model achieves superior performance compared to the baseline in simulated and real-world open-set speaker identification tasks.

\section{Household-adapted scoring model}
\vspace{-1em}
\begin{figure}[t]
\begin{minipage}[b]{1.0\linewidth}
  \centering
  \includegraphics[width=1.05\textwidth]{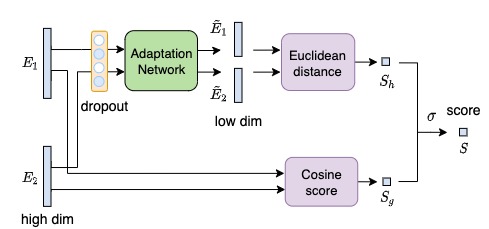}
\end{minipage}
\caption{Architecture of the household-adapted scoring model}
\label{fig:scoring_diagram}
\end{figure}

The household-adapted scoring back-end is shown in Figure \ref{fig:scoring_diagram}. For each household, we train the household-adapted scoring model to identify and discriminate among the speakers in the household, as well as guest speakers. The inputs to the scoring model are two utterance-level global embeddings $E_1, E_2 \in R^D$ {\color{black} ($D$ is the embedding dimension) extracted from corresponding audio features $X_1, X_2 \in R^{T \times F}$ ($T$ is the number of audio frames and $F$ is the audio feature dimension)} by the front-end speaker embedding extractor.
Global embeddings ($E_1, E_2$) are length-normalized.
The scoring model is designed to linearly fuse the global cosine score $S_g$ based on the global embedding space ($E_1, E_2$) with a Euclidean distance score $S_h$ based on a lower-dimensional space, where global embeddings are mapped into the household space by an adaptation network trained with household-specific pairs of data $\tilde{E}_1, \tilde{E}_2 \in R^K$ with $K < D$ ({\color{black}$K$ is the adapted lower dimension)}. Prior to the embedding adaptation network, we optionally used synchronous input vector dropout to generate masked inputs $E^*_1, E^*_2$. Then, we apply an affine transform into a lower dimension, followed by a ReLU non-linearity.
Note that the parameters of the embedding adaptation network ($\text{InputDropout}, W, B$) are shared for pairs of input embeddings $E_1, E_2$. {\color{black}$W, B$ are the learnable weights and biases of the adaptation network.}

The household-adapted Euclidean distance score $S_h$ aims to complement the global cosine score $S_g$ by focusing on the distribution of embeddings found in each household. {\color{black} We find empirically that the Euclidean distance at the low dimension space works better in our score fusion (see
Table~\ref{tab:ablation study}).} The final score is computed as a logistic regression with inputs $S_g$ and $S_h$, and lies between 0 and 1. The model parameters are trained in an end-to-end fashion via binary cross-entropy loss. In summary, our model performs the following computation: 
\begin{equation} 
\label{eq1}
\begin{split}
E_i &= \text{EmbeddingExtractor}(X_i),\ i=1,2 \\
S_g &= \text{CosineSimilarity}(E_1, E_2) \\
E^*_i &= \text{InputDropout}(E_i),\ i=1,2 \\
\tilde{E}_i &= \text{ReLU}(W E^*_i + B),\ i=1,2 \\
S_h &= \text{EuclideanDistance}(\tilde{E}_1, \tilde{E}_2) \\
S &= \text{Sigmoid}(w_1S_g +  w_2S_h + b) \\
\end{split}
\end{equation}
where $w_1,w_2,b$ are learnable weights and bias for the score fusion, and $S$ is the fused score based on which we identify or reject speakers.

For each household with $N$ enrolled speakers,
we construct positive and negative training pairs for contrastive learning (see Section \ref{sec:data_preparation} for details). Positive pairs ($\mathcal{S}_{pos}$) are formed from utterances of the same speaker in the household; negative pairs ($\mathcal{S}_{neg}$) are formed by utterances between household members and utterances between a household member and guest speakers from other households. To compensate for the greater number of negative examples  we use a fixed weight hyper-parameter $w = |\mathcal{S}_{neg}|/|\mathcal{S}_{pos}|$, in the loss function: 
\begin{equation}
   L = - \frac{1}{|\mathcal{S}_{pos}| + |\mathcal{S}_{neg}|} \left(w\sum_{i \in \mathcal{S}_{pos}} \log (S_i) + \sum_{i \in \mathcal{S}_{neg}} \log (1-S_i) \right)
\end{equation}
where {\color{black} $|\mathcal{S}_{neg}|$ and $|\mathcal{S}_{pos}|$ are the total number of positive and negative pairs, respectively;} $S_i$ is the fused score of a given pair ($S_i = \text{Score}(E_{i,1}, E_{i,2})$).

During inference, we precompute the average embeddings of each enrolled speaker ($E_{\text{enroll}}$).
For a household with $N$ enrolled speakers, we run the scoring of the test utterance against each enrolled speaker ($S^k = \text{Score}(E_{\text{enroll}}^k, E_{\text{test}}), \ k=1, \dots, N$) and identify the speaker with the maximum score: $S_{\text{max}} = \max(S^1, \dots, S^N)$.
If $S_{\text{max}}$ is greater than a predefined threshold, we accept the hypothesis that the test utterance is spoken by the speaker producing the maximum score (the rank-1 speaker). Otherwise we predict that the speaker is not enrolled (i.e. a guest).
{\color{black} This predefined threshold determines the trade-off between false accepts and rejections, and is selected based on application requirements.}

\begin{table*}[t]
	\caption{EER ($\%$) with household-adapted scoring model on random and hard households simulated with VoxCeleb1 data.
	$N$ is the number of speakers per household. 
	Numbers in parentheses show relative EER improvements over baseline. For the household-adapted scoring model, we use the household-adapted score only ($S_h$ {\bf only}) or both household-adapted score and global cosine score ($S_h$, $S_g$ {\bf both}, without or with dropout of 0.5) in logistic regression.}
	\label{tab:EER_voxcelb_1}
	\centering
	\small
	\begin{tabular}{ c | c l l l| c l l l}
	\toprule
	& \multicolumn{4}{c|}{\textbf{Random households}} & \multicolumn{4}{c}{\textbf{Hard households}}  \\
	\bf $N$ 
	& \bf Baseline &  \bf $S_h$ only & \bf $S_h, S_g$ both & \bf \begin{tabular}[c]{@{}l@{}}$S_h, S_g$ both \\ (dropout=0.5)\end{tabular}
	& \bf Baseline &  \bf $S_h$ only & \bf $S_h, S_g$ both & \bf \begin{tabular}[c]{@{}l@{}}$S_h, S_g$ both \\ (dropout=0.5)\end{tabular}\\
	\midrule
    2 & 2.66     & 1.87 (29.7\%)    & 1.63 (38.7\%) & \textbf{1.60 (39.8\%)} & 3.32     & 2.35 (29.2\%)    & 1.98 (40.4\%) & \textbf{1.82 (45.2\%)}\\
    3 & 3.40      & 2.59 (23.8\%)    & 2.14 (37.1\%) & \textbf{2.06 (39.4\%)} & 3.97     & 2.43 (38.8\%)    & 2.06 (48.1\%) & \textbf{1.70 (57.2\%)} \\
    4 & 3.85     & 3.14 (18.4\%)    & {2.57 (33.2\%)}  & \textbf{2.31 (40.0\%)} & 4.41     & 2.73 (38.1\%)    & {2.29 (48.1\%)}  & \textbf{1.65 (62.6\%)} \\
    5 & 3.98     & 3.81 (4.3\%)     & {2.83 (28.9\%)}  & \textbf{2.54 (36.2\%)}& 5.29     & 3.23 (38.9\%)    & {2.56 (51.6\%)}  & \textbf{1.54 (70.9\%)}\\
    6 & 4.58     & 4.68 (-2.2\%)    & {3.41 (25.5\%)} & \textbf{2.83 (38.2\%)} & 6.74     & 4.48 (33.5\%)    & {3.93 (41.7\%)}  & \textbf{2.78 (58.8\%)}\\
    7 & 4.86     & 5.58 (-14.8\%)   & {3.79 (22.0\%)}  & \textbf{2.97 (38.9\%)}& 7.87     & 5.67 (28.0\%)    & {4.34 (44.9\%)}   & \textbf{2.89 (62.3\%)}\\
    \bottomrule
    \end{tabular}
\end{table*}

\section{Experiments}
\vspace{-1em}
\subsection{Data preparation}\label{sec:data_preparation}
\noindent \textbf{Datasets.}
We used two distinct datasets to train and evaluate our back-end scoring models. Section \ref{sec:model_training} explains the front-end model training data. For back-end scoring model training and evaluation, first, we used VoxCeleb1 \cite{nagrani2017voxceleb} data to simulate households with 2-7 speakers each, with two different levels of difficulty, as detailed below. For each speaker, we sampled up to 50 utterances for training. For each household, we also sampled 250 guest utterances from speakers outside the household. We generated training pairs of utterances per households using a contrastive learning approach. Positive pairs are constructed by pairing up utterances from the same household member (e.g., A-A, B-B). Negative pairs are constructed by pairing up utterances between household members (e.g., A-B) as well as utterances of household members and randomly selected guest (G) utterances from other households (e.g., A-G and B-G). Positive and negative training pairs for each household were combined and shuffled before training. Training batch size was 1024. 

For a fully realistic evaluation, we used random de-identified customer utterances and household membership metadata from internal production data. We randomly selected over 1000 households with 2-4 speakers each.  We used the same method as on VoxCeleb1 to generate household-specific training pairs. Speaker labels were derived from enrollment, as well as by pseudo-labeling runtime utterances, as described below. 
\\

\noindent \textbf{VoxCeleb random household generation.} 
We constructed simulated households of different sizes to train and evaluate our household-adapted scoring model. For each household size ranging from 2 to 7 speakers, we synthesized 1000 households. 
Although typical multi-speaker households have between 2 and 5 speakers, we simulated households with up to 7 speakers for model evaluation. For each household of size $N$, we randomly sample $N$ speakers without replacement from the 1251 speakers in VoxCeleb1.
A speaker can appear in different simulated households with different fellow members.
For each speaker, 4 utterances are used as enrollment utterances, 10 utterances are used as evaluation utterances, and up to 50 (due to limited number of utterances for some speakers) are used for training. For each utterance, we randomly crop a 2-second segment and discard other parts in order to mimic the typical amount of speech in smart speaker user utterances.
We averaged length-normalized global embeddings for enrollment utterances to produce speaker profiles.
Global embeddings were extracted by a CNN model (see Section~\ref{sec:model_training}).
\\

\noindent \textbf{VoxCeleb hard household generation.} 
We further constructed ``hard'' simulated households of different sizes by pairing highly similar (i.e., confusable) speakers.
Two speakers are considered highly similar if the cosine similarity of their speaker-level embeddings is larger than a threshold. Moreover, to construct a hard household with more than two speakers, we require every two speakers in the household to be highly similar.  Specifically, speaker-level embeddings are constructed by averaging utterance-level embeddings from 20 randomly selected utterances (using 2-second snippets). Utterance-level embeddings are extracted by a CNN model (see Section \ref{sec:model_training}). The cosine similarity threshold for highly similar speaker pairs is chosen as 0.13, which is in the  98th percentile of all cosine similarity scores between utterances of different speakers.
\\

\noindent \textbf{VoxCeleb training utterance label corruption.} 
In real-world applications, labels of utterances collected from smart home devices are typically generated by pseudo-labeling methods based on semi-supervised learning. In general, the pseudo-labeling process produces label errors and thus causes performance degradation. To evaluate the robustness of our approach to label noise, we randomly corrupted data labels for training utterances (up to 50 utterances per speaker) and trained and evaluated our models on the corrupted data. Specifically, we assume at first the training utterances from all speakers in a household are unlabeled. Given an unlabeled utterance and a corruption error rate $\epsilon$, we annotated the utterance with a ground truth label with probability $1-\epsilon$, and annotate it with a random speaker label within the household otherwise. Here the label corruption mimics a pseudo-labeled error rate of~$\epsilon$, which we varied from $0.0$ to $0.1$ in our robustness study.
\\
 
\noindent \textbf{Internal data training utterance pseudo-labeling.}
We used a text-independent speaker verification model (see Section \ref{sec:model_training}) to generate embeddings for experiments on internal production data. To generate high quality training data with few pseudo-label errors, we selected runtime utterances whose rank-1 score is higher than a predefined threshold and where the difference between rank-1 and rank-2 scores is above a second predefined threshold. 

\vspace{-1em}
\subsection{Model training}\label{sec:model_training}
\noindent \textbf{Front-end embedding extractor.} The CNN model used as the front-end embedding extractor to train the household-adapted scoring model with VoxCeleb1 data is Half-ResNet34 \cite{chung2020defence,heo2020clova}, which has half of the channel numbers of the original ResNet34. 
The model was trained on VoxCeleb2 \cite{chung2018voxceleb2} with 5994 speakers for 100 epochs on a single Tesla V100 Nvidia GPU. The training used additive margin softmax loss \cite{wang2018cosface} with a margin of 0.1. We used the Adam optimizer with the learning rate initially at $10^{-3}$ for 5000 steps and then decaying every 2500 steps with a factor of 0.9. The output embedding dimension is 256. 

The front-end embedding extractor to train the household-adapted scoring model with internal production data was implemented as a text-independent model with three-layer LSTM with 512 hidden units per layer \cite{wan2018generalized}. The model was trained using de-identified utterances from internal production data with metric learning objectives. The output embedding dimension is 512.
\\

\noindent \textbf{Household-adapted back-end scoring model.} 
We used one fully connected layer with 32 neurons and ReLU activation function as the adaptation network in the household-adapted scoring model, which can be interpreted as performing an affine transform  towards a lower dimension with non-linearity transformation. Each household-adapted scoring model was trained using the household-specific training data from VoxCeleb1 for 10 epochs with a learning rate of 0.01. On internal production data, validation error stabilized after training for 3-5 epochs. 
\\

\noindent \textbf{Input dropout for global embedding.}
In experiments with dropout, we applied a dropout layer to the global embeddings input to the adaption network (see Figure \ref{fig:scoring_diagram}). For each pair of input vectors, the same randomly chosen vector component was masked out in the two paired embeddings.
\\
\vspace{-2em}
\subsection{Evaluation}
False accept rate (FAR) is the fraction of guest utterances that are falsely accepted as enrolled speakers. False negative identification rate (FNIR) is the fraction of enrolled speaker utterances that are misidentified as a different speaker or rejected as guests. Equal error rate (EER) is defined as the value where FAR is equal to FNIR. We used EER to evaluate speaker identification performance independent of the priors for positive/negative test cases. The baseline is based only on the cosine score, which is the cosine similarity scaled to the interval $[0,1]$.

\vspace{-1em}
\section{Results}
\vspace{-1em}
\subsection{Household-adapted scoring with VoxCeleb data}
First, we evaluated the household-adapted scoring model on \textit{random} simulated  households. {\color{black} The VoxCeleb utterances could be recorded in different acoustic environments via different devices. However, such channel effects are expected to be minimal for utterances recorded on shared devices in the same household. We mitigated this mismatch by considering \textit{hard} simulated households and internal real-life data, as described below.} 
As shown in Table \ref{tab:EER_voxcelb_1}, EER of the baseline increases with household size, which is expected given that speaker identification becomes harder when there are more enrolled speakers. 
Notably, when using only the household-adapted score ($S_h$ {\bf only} in Table~\ref{tab:EER_voxcelb_1}), EER improves for smaller households  ($N = 2,\ldots,5$), while it degrades for larger households ($N = 6,7$) {\color{black} with respect to the baseline}. When fusing the household-adapted score and the global cosine score ($S_h$, $S_g$ {\bf both} in Table~\ref{tab:EER_voxcelb_1}), we observe relative EER reduction of between 22\% and 39\% compared to the baseline. 
This result demonstrates the benefit of fusing household-adapted score and global cosine score. We hypothesize that the fused score helps to counteract overfitting of the household-adapted scoring model, as well as to better identify guest speaker utterances, which are unseen during adaptation training.

Next, we evaluated the model performance on \textit{hard} households. Simulating hard households better mimics the real-life fact that people living together tend to share similar acoustic environments, accents, or other voice features.
As shown in Table~\ref{tab:EER_voxcelb_1}, keeping the model constant, baseline EER for hard households is higher than {\color{black} for} random households {\color{black}of the corresponding size}, as intended.
When using only the household-adapted score ($S_h$ {\bf only} in Table~\ref{tab:EER_voxcelb_1}), we observe relative EER improvement of 28\% to 39\% depending on household size. 
When fusing household-adapted and global cosine scores ($S_h$, $S_g$ {\bf both} in Table~\ref{tab:EER_voxcelb_1}), the relative EER improvement goes up further, to between 40\% and 52\%. This result shows that our household-adapted scoring model performs well even in challenging scenarios, showing more benefit in hard households than in random households. 

As a control experiment,
we trained a shared scoring model with the same architecture as the household-adapted model (Figure~\ref{fig:scoring_diagram}) using VoxCeleb1 data from {\em all} households of a given size $N$. The shared scoring model performed worse than either the baseline or the household-adapted model, likely because the small embedding adaptation layer does not have the capacity to discriminate between speakers from all households.
This demonstrates that the benefits we observed from the household-adapted scoring model are not due simply to the additional training data or the additional model parameters.

\begin{table}[t]
	\caption{Ablation studies using hard four-speaker households. Baseline EER is 4.41\%. In each row, we show EER and relative EER improvement ($\Delta$) compared to baseline.}
	\label{tab:ablation study}
	\centering
	\footnotesize
	\begin{tabular}{c| c c c c|c c}
	\toprule
	            & $S_g$ & $S_h$ & \bf \begin{tabular}[c]{@{}c@{}}Adaptation\\ layers \end{tabular} & \bf Dropout & \bf EER & \bf Rel. $\Delta$\\
    \midrule
    1 & yes     & Euclidean     & one (32)         & 0            &  2.29 &  48.1\% \\
    2 & \bf no  & Euclidean     & one (32)         & 0            &  2.73 &  38.1\% \\
    3 & yes     & \bf cosine    & one (32)         & 0            &  5.31 & -20.4\% \\
    4 & yes     & Euclidean     & \bf two (64, 32) & 0            &  2.72 &  38.3\% \\
    5 & yes     & Euclidean     & one (32)         & \bf 0.2      &  1.79 &  59.4\% \\
    6 & yes     & Euclidean     & one (32)         & \bf 0.5      &  \bf 1.65 &  \bf  62.6\% \\
    7 & yes     & Euclidean     & one (32)         & \bf 0.8      &  2.58 &  41.5\% \\
    \bottomrule
    \end{tabular}
\end{table}

\vspace{-1em}
\subsection{Ablation studies}
We conducted ablation studies to optimize the architecture of the household-adapted scoring system (Table \ref{tab:ablation study}). As shown in the previous section, dropping global cosine score ($S_g$) hurts performance (row 2 in Table \ref{tab:ablation study}) {\color{black} compared to using $S_g$ and $S_h$ score fusion (row 1 in Table \ref{tab:ablation study})}. Using Euclidean distance on the low dimensional embedding outperforms cosine similarity score (row 3 in Table \ref{tab:ablation study}). Increasing the number of layers in the adaptation network does not improve performance (row~4 in Table~\ref{tab:ablation study}). Interestingly, adding dropout before the adaptation network significantly improves performance (rows~5-7 in Table~\ref{tab:ablation study}). Among various dropout rates, 0.5 gave the best results. 
 
To further evaluate the benefit of dropout on the performance, we conducted experiments for household-adapted models with dropout rate 0.5 for all random and hard households. As shown in Table~\ref{tab:EER_voxcelb_1}, models with dropout outperformed models without dropout and reduced EER by 36\% to 40\% for random households, and by 45\% to 71\% for hard households, compared to baseline.
We hypothesize that input-side dropout helps against overfitting to household-specific data and keeps more useful information for identifying speakers in a smaller-dimensional household-adapted embedding space.
{\color{black} Interestingly, we observe that $N=5$ has the lowest EER. We hypothesize that the optimal dropout rate for different household sizes could vary.}

\begin{figure}[tb]
\begin{minipage}[b]{1.0\linewidth}
  \centering
  \centerline{\includegraphics[width=0.9\textwidth]{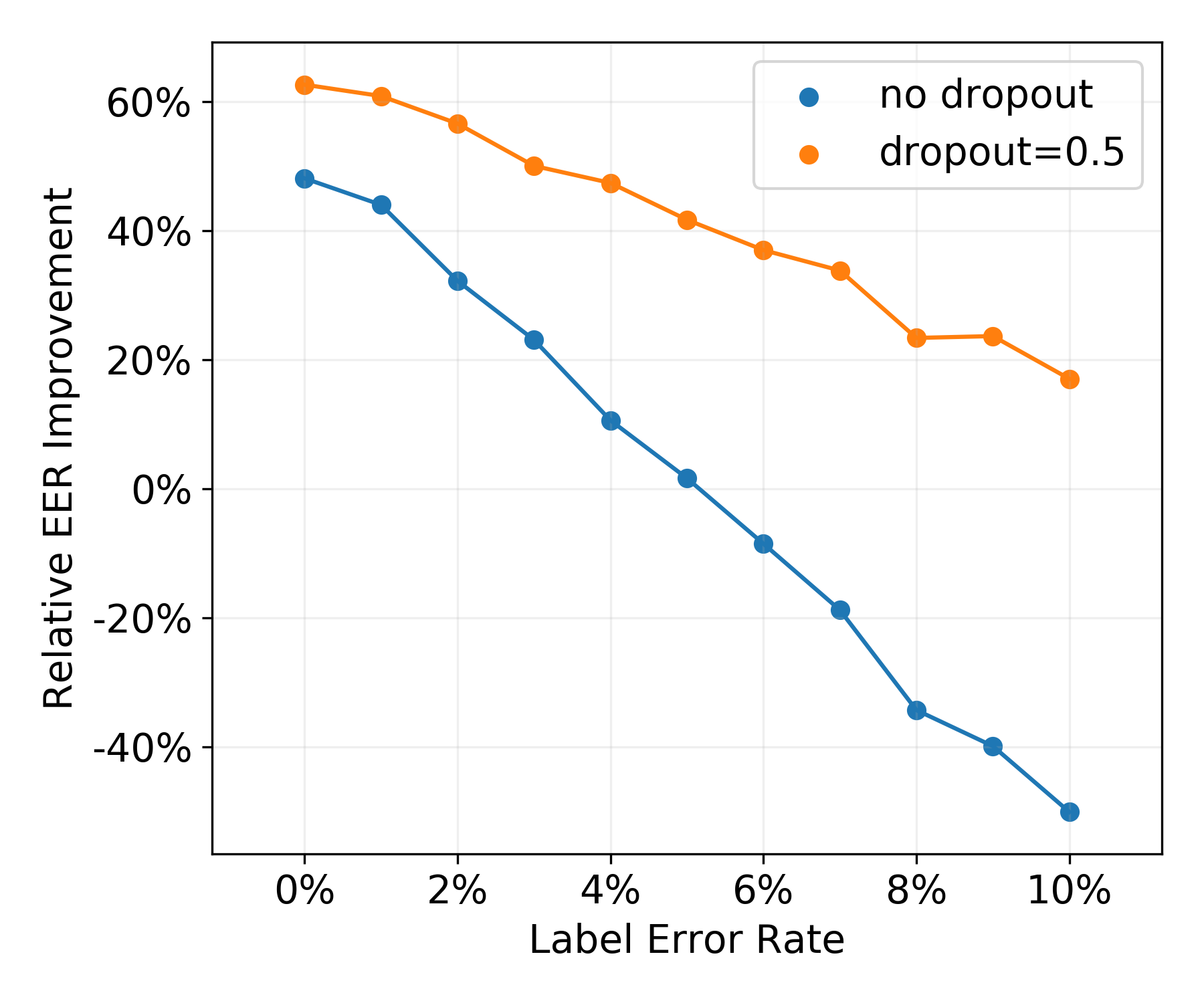}}
\end{minipage}
\caption{Impact of training data labeling error rate on performance of the household-adapted scoring. We simulated four-speaker \textit{hard} households using VoxCeleb1 data and randomly corrupted speaker labels with a varying label error rate.}
\label{fig:label_error}
\end{figure}
\vspace{-1em}
\subsection{Impact of training data pseudo-label error} 
To evaluate the impact of mislabeling in training data, we trained and evaluated our models on VoxCeleb1 synthesized households with randomly corrupted speaker labels (see Section \ref{sec:data_preparation}). Figure~\ref{fig:label_error} shows evaluation result at different label error rates. The performance of household-adapted scoring without dropout degrades quickly as the label error rate increases. At around 5\% label error, the benefit of household-adapted scoring model is offset by the poor label quality. However, with dropout, the performance gain degrades more slowly, such that even with a 10\% error rate there is still about 17\% reduction in EER.  Input dropout effectively mitigates the impact of labeling noise.
{\color{black} Pseudo-labeling error rates in practice typically fall into the 0-10\% range.} Additionally, advanced pseudo-labeling strategies such as label propagation \cite{berthelot2019mixmatch, xie2019unsupervised, chen2021graph} or teacher models \cite{lee2013pseudo, pham2021meta} can be further employed to reduce the pseudo-label error rate, which would benefit the household scoring model in particular.
\vspace{-1em}
\subsection{Household-adapted embedding visualization}
\begin{figure}[tb]
\begin{subfigure}{.23\textwidth}
  \centering
  \includegraphics[width=\linewidth]{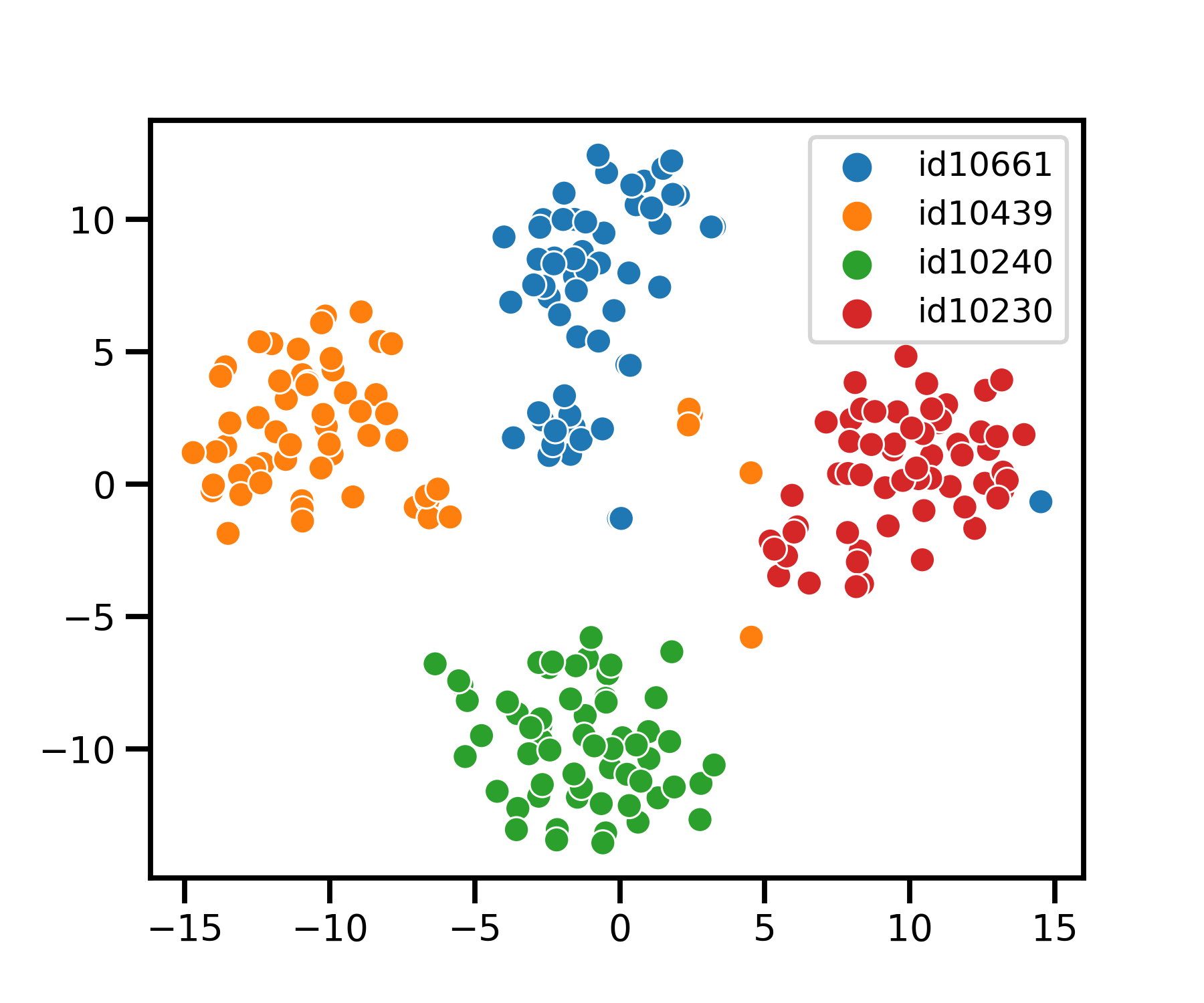}
  \caption{Original ($N$=4)}
  \label{fig:sfig1}
\end{subfigure}%
\begin{subfigure}{.23\textwidth}
  \centering
  \includegraphics[width=\linewidth]{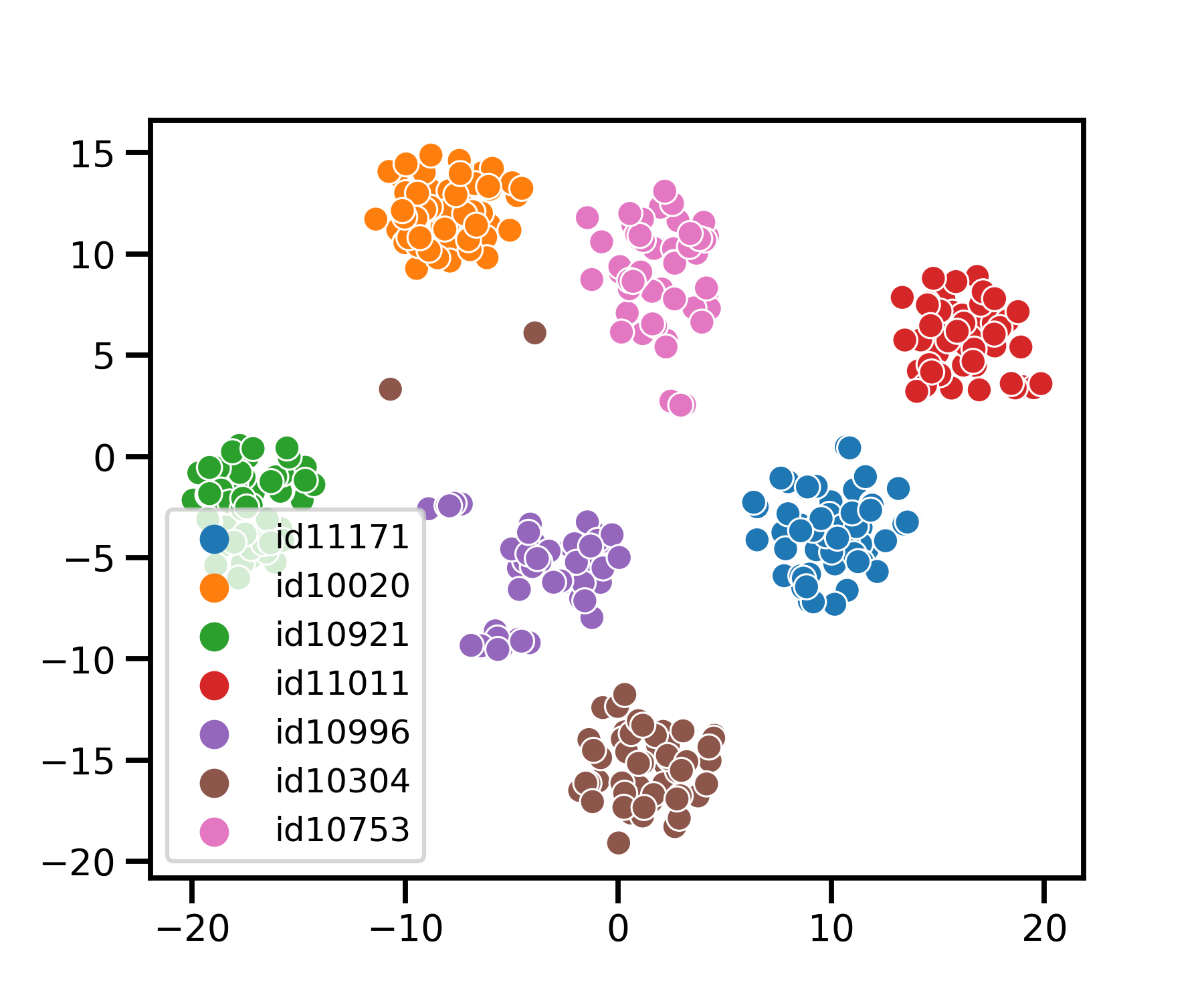}
  \caption{Original ($N$=7)}
  \label{fig:sfig2}
\end{subfigure}
\begin{subfigure}{.23\textwidth}
  \centering
  \includegraphics[width=\linewidth]{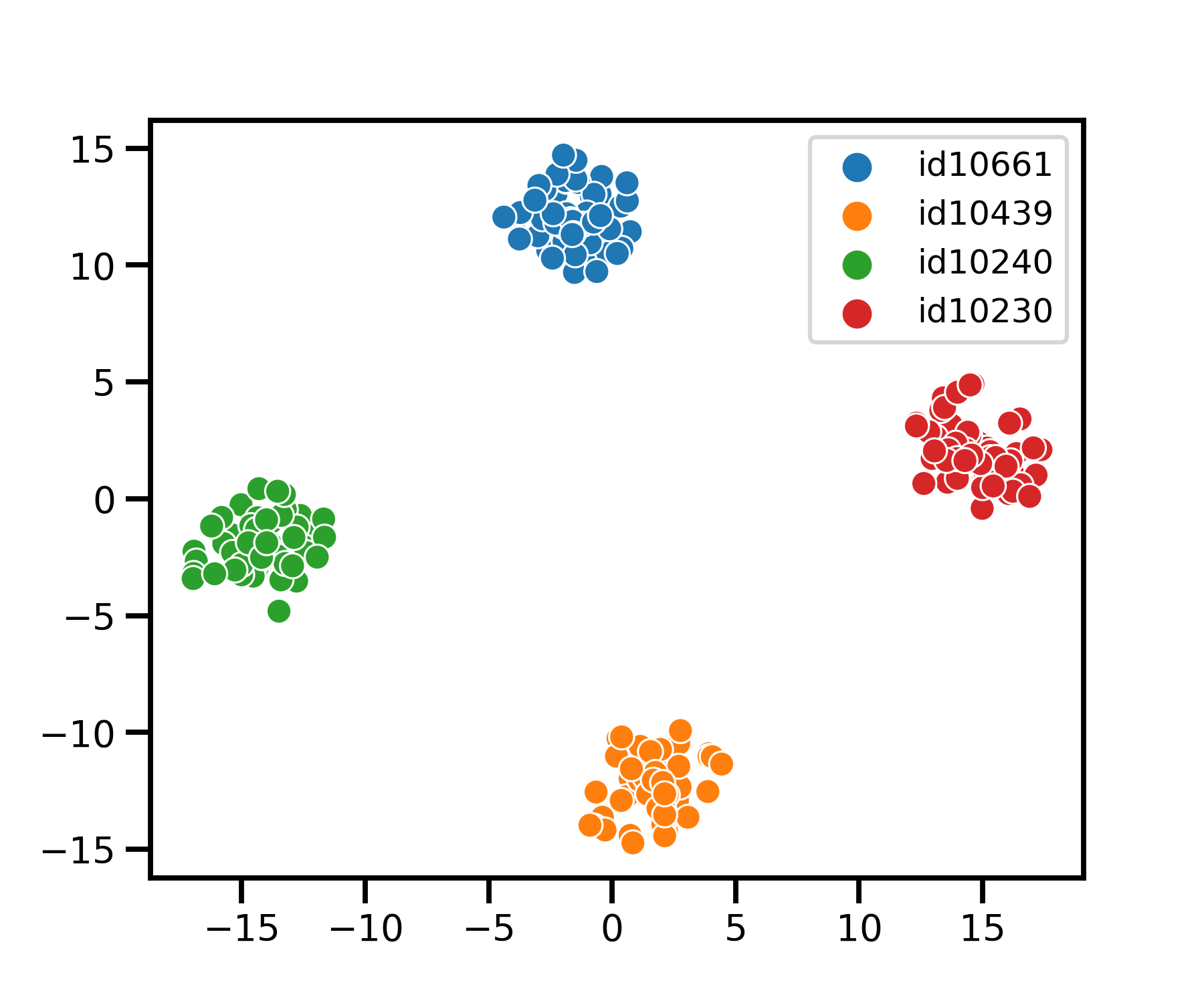}
  \caption{No dropout ($N$=4)}
  \label{fig:sfig3}
\end{subfigure}
\begin{subfigure}{.23\textwidth}
  \centering
  \includegraphics[width=\linewidth]{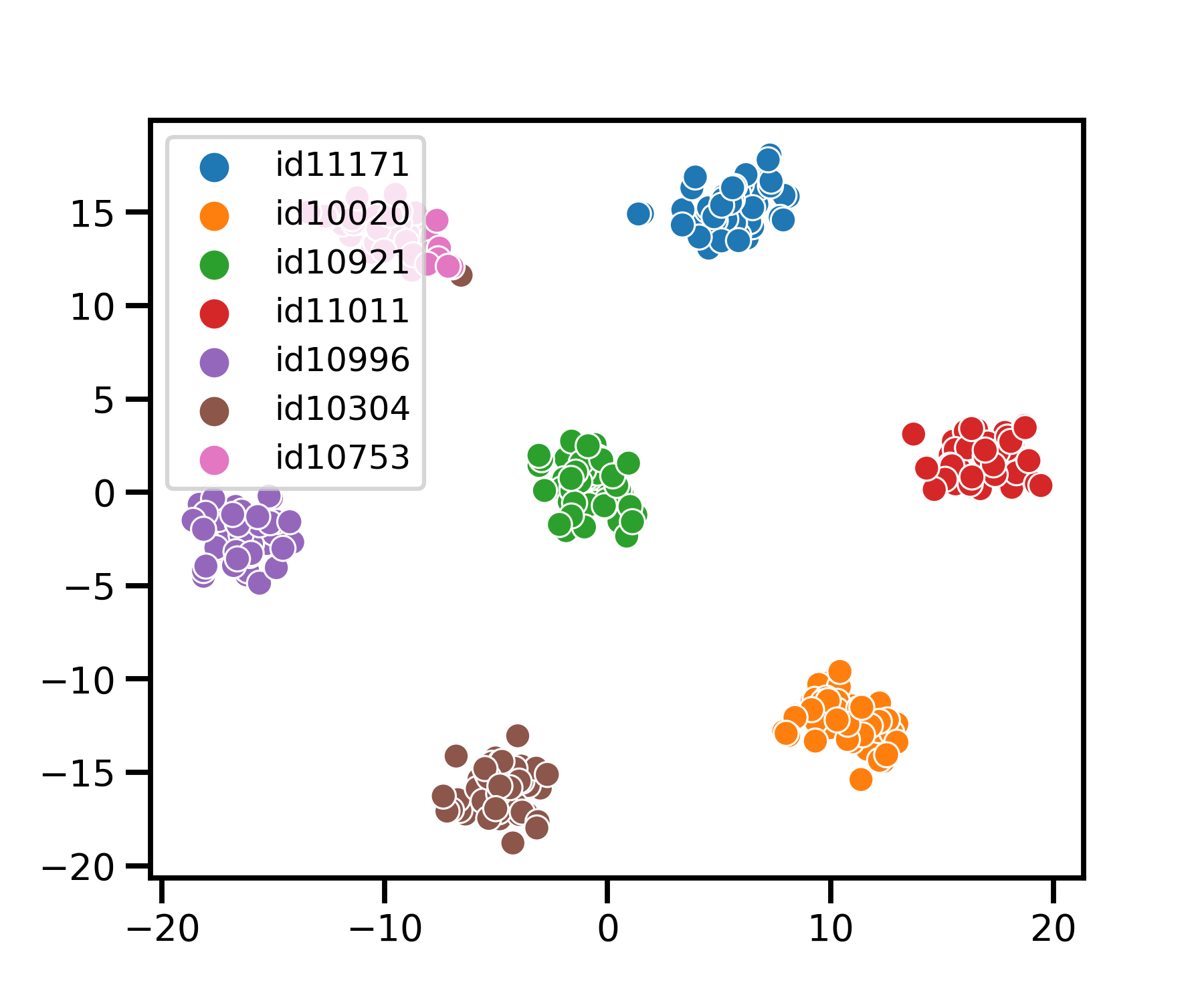}
  \caption{No dropout ($N$=7)}
  \label{fig:sfig4}
\end{subfigure}

\begin{subfigure}{.23\textwidth}
  \centering
  \includegraphics[width=\linewidth]{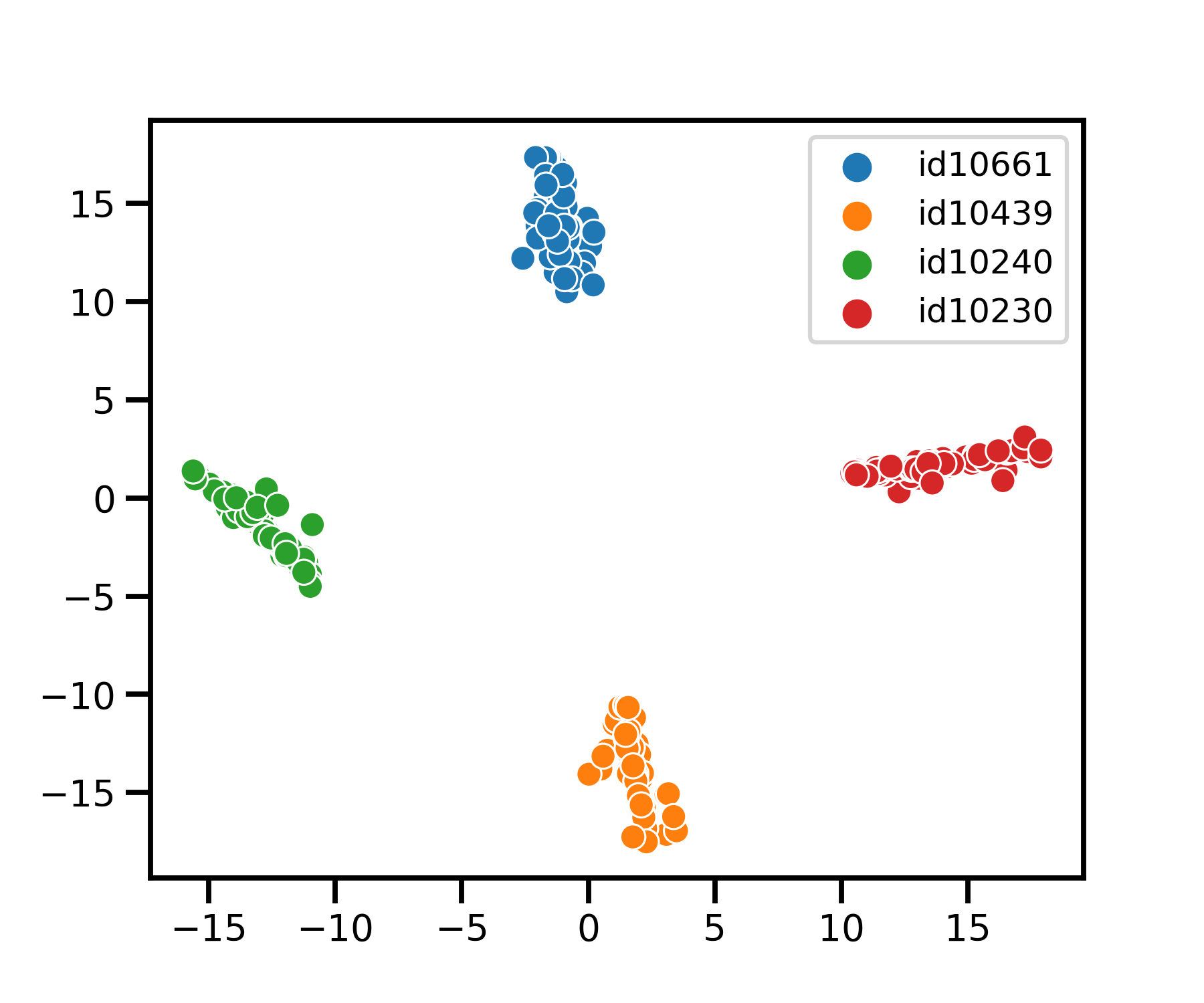}
  \caption{Dropout=0.5 ($N$=4)}
  \label{fig:sfig5}
\end{subfigure}
\begin{subfigure}{.23\textwidth}
  \centering
  \includegraphics[width=\linewidth]{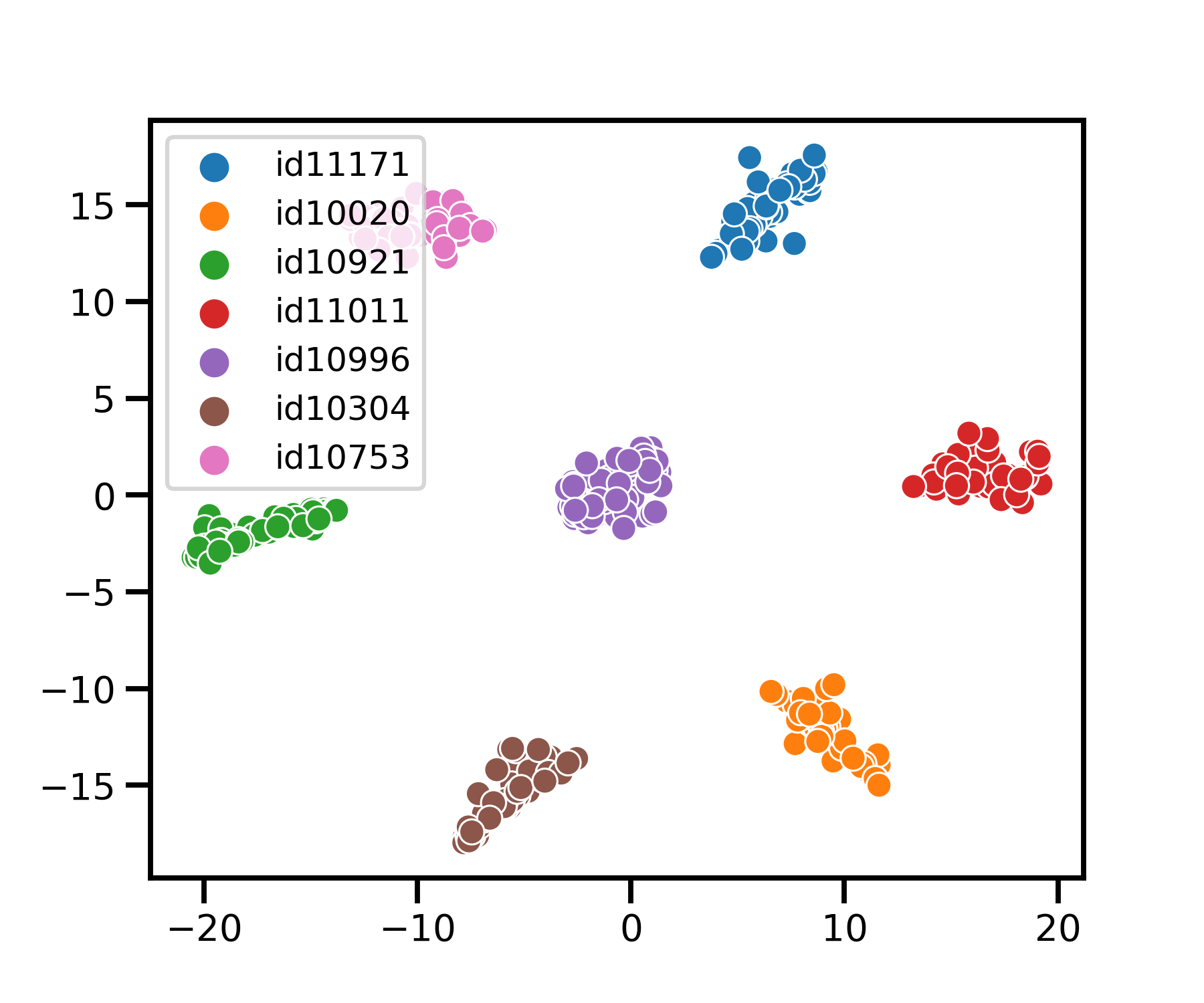}
  \caption{Dropout=0.5 ($N$=7)}
  \label{fig:sfig6}
\end{subfigure}

\caption{The t-SNE visualization of original embeddings (a, b) and the household-adapted embeddings projected by adaptation network in the household-adapted scoring model (c, d: no dropout, e, f: dropout=0.5). A four-speaker hard household (a, c, e) and a seven-speaker hard household (b, d, f) are selected at random. Utterances in both training and evaluation data are included for visualization. Euclidean distance was used as a distance metric in t-SNE.}
\label{fig:tSNE}
\end{figure}

To understand the performance gain from household-adapted scoring, we used t-SNE to visualize the embeddings from the front-end embedding extractor as well as the household adapted embeddings from the adaptation network. As shown in Figure \ref{fig:tSNE}, the speaker clusters in the household-adapted embedding space are much more compact, evenly distributed, and better separated. Between-speaker separation is increased and within-speaker variability is decreased. We hypothesize that the adaptation network learns a household-adapted embedding subspace in which speaker embeddings become more uniform and aligned \cite{wang2020understanding}.
Remarkably, with the usage of input-side dropout, the adapted embedding clusters are even more compact. They show larger margins between each other, thus reflecting the additional performance gain.

\vspace{-1em}

\subsection{Application to internal data}
To validate our household-adapted scoring model on data involving speakers from real-life households and with far-field audio, we trained and evaluated on internal data, as described in Section~\ref{sec:data_preparation}.
Furthermore, this evaluation relied on realistic pseudo-labeling in the training data.
The household-adapted scoring model reduced EER by 42.5\% (without dropout) and 49.2\% (with dropout rate 0.5) compared to baseline cosine scoring. These improvements are broadly consistent with those observed on the simulated households based on VoxCeleb data, giving us further confidence in the approach.

\section{Conclusions}
\vspace{-1em}
We have developed a household-adapted scoring model for speaker identification through household embedding adaptation together with global and local score fusion. Our approach achieves significant error reduction in open-set speaker identification problems for small subsets of speaker, where the original speaker embeddings are trained on large speaker sets, making the approach ideal for household and on-device applications.  The household-adapted scoring model not only improves speaker identification on synthetic random households (relative EER improvements of 36\% to 40\%), but also on households with confusable speakers (relative EER improvement 45\% to 71\%). Moreover, we showed that household-adapted scoring is robust to label noise, mainly as a result of input-side dropout in adaptation. With pseudo-labeled training data, for four-speaker hard households, we observed a relative EER improvement of 17\% even with 10\% labeling error. Finally, we demonstrated that the household-adapted scoring model improved real-world speaker identification on internal data (relative EER improvement of 49.2\%).

Our embedding adaptation approach can be applied not only to households, but to other scenarios where speaker recognition systems need to be tuned to speaker subsets. For example, we can use it to differentiate among meeting participants who all share a regional dialect. In future work, we plan to investigate extensions of the model to very small adaptation sets and in the absence of pseudo-labeled data.

\section{Acknowledgments}
\vspace{-1em}
We wish to thank our colleagues (in alphabetical order) Long Chen, Zeya Chen, Roger Cheng, Daniel Garcia-Romero, Minho Jin, Chelsea J.-T. Ju, Ryan Langman, Ruirui Li, Hongda Mao, Yixiong Meng, and Guoli Sun, for their valuable discussion and feedback, and managers Oguz Elibol, Mohamed El-Geish and Itay Teller, for their input and support. 

\normalsize
\bibliographystyle{IEEEtran}
\bibliography{refs}

\begin{thebibliography}{10}
\providecommand{\url}[1]{#1}
\csname url@samestyle\endcsname
\providecommand{\newblock}{\relax}
\providecommand{\bibinfo}[2]{#2}
\providecommand{\BIBentrySTDinterwordspacing}{\spaceskip=0pt\relax}
\providecommand{\BIBentryALTinterwordstretchfactor}{4}
\providecommand{\BIBentryALTinterwordspacing}{\spaceskip=\fontdimen2\font plus
\BIBentryALTinterwordstretchfactor\fontdimen3\font minus
  \fontdimen4\font\relax}
\providecommand{\BIBforeignlanguage}[2]{{%
\expandafter\ifx\csname l@#1\endcsname\relax
\typeout{** WARNING: IEEEtran.bst: No hyphenation pattern has been}%
\typeout{** loaded for the language `#1'. Using the pattern for}%
\typeout{** the default language instead.}%
\else
\language=\csname l@#1\endcsname
\fi
#2}}
\providecommand{\BIBdecl}{\relax}
\BIBdecl

\bibitem{Li2020}
\BIBentryALTinterwordspacing
R.~Li, J.-Y. Jiang, X.~Wu, C.-C. Hsieh, and A.~Stolcke, ``{Speaker
  Identification for Household Scenarios with Self-Attention and Adversarial
  Training},'' in \emph{Proc. Interspeech 2020}, 2020, pp. 2272--2276.
  [Online]. Available: \url{http://dx.doi.org/10.21437/Interspeech.2020-3025}
\BIBentrySTDinterwordspacing

\bibitem{fortuna2005open}
J.~Fortuna, P.~Sivakumaran, A.~Ariyaeeinia, and A.~Malegaonkar, ``Open-set
  speaker identification using adapted gaussian mixture models,'' in
  \emph{Ninth European Conference on Speech Communication and Technology},
  2005.

\bibitem{angkititrakul2007discriminative}
P.~Angkititrakul and J.~H. Hansen, ``Discriminative in-set/out-of-set speaker
  recognition,'' \emph{IEEE Transactions on Audio, Speech, and Language
  Processing}, vol.~15, no.~2, pp. 498--508, 2007.

\bibitem{wilkinghoff2020open}
K.~Wilkinghoff, ``On open-set speaker identification with ivectors,'' in
  \emph{The Speaker and Language Recognition Workshop (Odyssey). ISCA}, 2020,
  pp. 408--414.

\bibitem{dehak2010front}
N.~Dehak, P.~J. Kenny, R.~Dehak, P.~Dumouchel, and P.~Ouellet, ``Front-end
  factor analysis for speaker verification,'' \emph{IEEE Transactions on Audio,
  Speech, and Language Processing}, vol.~19, no.~4, pp. 788--798, 2010.

\bibitem{heigold2016end}
G.~Heigold, I.~Moreno, S.~Bengio, and N.~Shazeer, ``End-to-end text-dependent
  speaker verification,'' in \emph{2016 IEEE International Conference on
  Acoustics, Speech and Signal Processing (ICASSP)}.\hskip 1em plus 0.5em minus
  0.4em\relax IEEE, 2016, pp. 5115--5119.

\bibitem{wan2018generalized}
L.~Wan, Q.~Wang, A.~Papir, and I.~L. Moreno, ``Generalized end-to-end loss for
  speaker verification,'' in \emph{Proc.\ IEEE ICASSP}, 2018, pp. 4879--4883.

\bibitem{chung2018voxceleb2}
J.~S. Chung, A.~Nagrani, and A.~Zisserman, ``Voxceleb2: Deep speaker
  recognition,'' \emph{arXiv preprint arXiv:1806.05622}, 2018.

\bibitem{snyder2017deep}
D.~Snyder, D.~Garcia-Romero, D.~Povey, and S.~Khudanpur, ``Deep neural network
  embeddings for text-independent speaker verification.'' in
  \emph{Interspeech}, 2017, pp. 999--1003.

\bibitem{snyder2018x}
D.~Snyder, D.~Garcia-Romero, G.~Sell, D.~Povey, and S.~Khudanpur,
  ``{X-vectors}: Robust {DNN} embeddings for speaker recognition,'' in
  \emph{Proc.\ IEEE ICASSP}, 2018, pp. 5329--5333.

\bibitem{li2017deep}
C.~Li, X.~Ma, B.~Jiang, X.~Li, X.~Zhang, X.~Liu, Y.~Cao, A.~Kannan, and Z.~Zhu,
  ``Deep speaker: an end-to-end neural speaker embedding system,'' \emph{arXiv
  preprint arXiv:1705.02304}, vol. 650, 2017.

\bibitem{garcia2020magneto}
D.~Garcia-Romero, G.~Sell, and A.~McCree, ``Magneto: X-vector magnitude
  estimation network plus offset for improved speaker recognition,'' in
  \emph{Proc. Odyssey 2020 The Speaker and Language Recognition Workshop},
  2020, pp. 1--8.

\bibitem{desplanques2020ecapa}
B.~Desplanques, J.~Thienpondt, and K.~Demuynck, ``Ecapa-tdnn: Emphasized
  channel attention, propagation and aggregation in tdnn based speaker
  verification,'' \emph{arXiv preprint arXiv:2005.07143}, 2020.

\bibitem{ioffe2006probabilistic}
S.~Ioffe, ``Probabilistic linear discriminant analysis,'' in \emph{European
  Conference on Computer Vision}.\hskip 1em plus 0.5em minus 0.4em\relax
  Springer, 2006, pp. 531--542.

\bibitem{prince2007probabilistic}
S.~J. Prince and J.~H. Elder, ``Probabilistic linear discriminant analysis for
  inferences about identity,'' in \emph{2007 IEEE 11th International Conference
  on Computer Vision}.\hskip 1em plus 0.5em minus 0.4em\relax IEEE, 2007, pp.
  1--8.

\bibitem{kenny2010bayesian}
P.~Kenny, ``Bayesian speaker verification with heavy-tailed priors.'' in
  \emph{Odyssey}, vol.~14, 2010.

\bibitem{ferrer2020speaker}
L.~Ferrer and M.~McLaren, ``A speaker verification backend for improved
  calibration performance across varying conditions,'' \emph{arXiv preprint
  arXiv:2002.03802}, 2020.

\bibitem{pelecanos2021dr}
J.~Pelecanos, Q.~Wang, and I.~L. Moreno, ``Dr-vectors: Decision residual
  networks and an improved loss for speaker recognition,'' \emph{arXiv preprint
  arXiv:2104.01989}, 2021.

\bibitem{ye2020few}
H.-J. Ye, H.~Hu, D.-C. Zhan, and F.~Sha, ``Few-shot learning via embedding
  adaptation with set-to-set functions,'' in \emph{Proceedings of the IEEE/CVF
  Conference on Computer Vision and Pattern Recognition}, 2020, pp. 8808--8817.

\bibitem{shum2011exploiting}
S.~Shum, N.~Dehak, E.~Chuangsuwanich, D.~Reynolds, and J.~Glass, ``Exploiting
  intra-conversation variability for speaker diarization,'' in \emph{Twelfth
  Annual Conference of the International Speech Communication Association},
  2011.

\bibitem{konevcny2016federated}
J.~Kone{\v{c}}n{\`y}, H.~B. McMahan, F.~X. Yu, P.~Richt{\'a}rik, A.~T. Suresh,
  and D.~Bacon, ``Federated learning: Strategies for improving communication
  efficiency,'' \emph{arXiv preprint arXiv:1610.05492}, 2016.

\bibitem{nagrani2017voxceleb}
A.~Nagrani, J.~S. Chung, and A.~Zisserman, ``Voxceleb: a large-scale speaker
  identification dataset,'' \emph{arXiv preprint arXiv:1706.08612}, 2017.

\bibitem{chung2020defence}
J.~S. Chung, J.~Huh, S.~Mun, M.~Lee, H.~S. Heo, S.~Choe, C.~Ham, S.~Jung, B.-J.
  Lee, and I.~Han, ``In defence of metric learning for speaker recognition,''
  \emph{arXiv preprint arXiv:2003.11982}, 2020.

\bibitem{heo2020clova}
H.~S. Heo, B.-J. Lee, J.~Huh, and J.~S. Chung, ``Clova baseline system for the
  voxceleb speaker recognition challenge 2020,'' \emph{arXiv preprint
  arXiv:2009.14153}, 2020.

\bibitem{wang2018cosface}
H.~Wang, Y.~Wang, Z.~Zhou, X.~Ji, D.~Gong, J.~Zhou, Z.~Li, and W.~Liu,
  ``Cosface: Large margin cosine loss for deep face recognition,'' in
  \emph{Proceedings of the IEEE conference on computer vision and pattern
  recognition}, 2018, pp. 5265--5274.

\bibitem{berthelot2019mixmatch}
D.~Berthelot, N.~Carlini, I.~Goodfellow, N.~Papernot, A.~Oliver, and C.~Raffel,
  ``Mixmatch: A holistic approach to semi-supervised learning,'' \emph{arXiv
  preprint arXiv:1905.02249}, 2019.

\bibitem{xie2019unsupervised}
Q.~Xie, Z.~Dai, E.~Hovy, M.-T. Luong, and Q.~V. Le, ``Unsupervised data
  augmentation for consistency training,'' \emph{arXiv preprint
  arXiv:1904.12848}, 2019.

\bibitem{chen2021graph}
L.~Chen, V.~Ravichandran, and A.~Stolcke, ``Graph-based label propagation for
  semi-supervised speaker identification,'' \emph{arXiv preprint
  arXiv:2106.08207}, 2021.

\bibitem{lee2013pseudo}
D.-H. Lee \emph{et~al.}, ``Pseudo-label: The simple and efficient
  semi-supervised learning method for deep neural networks,'' in \emph{Workshop
  on challenges in representation learning, ICML}, vol.~3, no.~2, 2013.

\bibitem{pham2021meta}
H.~Pham, Z.~Dai, Q.~Xie, and Q.~V. Le, ``Meta pseudo labels,'' in
  \emph{Proceedings of the IEEE/CVF Conference on Computer Vision and Pattern
  Recognition}, 2021, pp. 11\,557--11\,568.

\bibitem{wang2020understanding}
T.~Wang and P.~Isola, ``Understanding contrastive representation learning
  through alignment and uniformity on the hypersphere,'' in \emph{International
  Conference on Machine Learning}.\hskip 1em plus 0.5em minus 0.4em\relax PMLR,
  2020, pp. 9929--9939.

\end{thebibliography}

\end{document}